# Wireless, Customizable Coaxially-shielded Coils for Magnetic Resonance Imaging

Ke Wu[1,3†], Xia Zhu[1,3†], Stephan W. Anderson[2,3*], Xin Zhang[1,3*]

[1]Department of Mechanical Engineering, Boston University, Boston, MA 02215, United States.

[2]Boston University Chobanian & Avedisian School of Medicine, Boston, MA, 02118, United States.

[3]Photonics Center, Boston University, Boston, MA 02215, USA.

[†]These authors contributed equally to this work.

*Corresponding author. E-mail: xinz@bu.edu (X.Z.); sande@bu.edu (S.W.A.)

## Abstract

Anatomy-specific RF receive coil arrays routinely adopted in magnetic resonance imaging (MRI) for signal acquisition, are commonly burdened by their bulky, fixed, and rigid configurations, which may impose patient discomfort, bothersome positioning, and suboptimal sensitivity in certain situations. Herein, leveraging coaxial cables' inherent flexibility and electric field confining property, for the first time, we present wireless, ultra-lightweight, coaxially-shielded MRI coils achieving a signal-to-noise ratio (SNR) comparable to or surpassing that of commercially available cutting-edge receive coil arrays with the potential for improved patient comfort, ease of implementation, and significantly reduced costs. The proposed coils demonstrate versatility by functioning both independently in form-fitting configurations, closely adapting to relatively small anatomical sites, and collectively by inductively coupling together as metamaterials, allowing for extension of the field-of-view of their coverage to encompass larger anatomical regions without compromising coil sensitivity. The wireless, coaxially-shielded MRI coils reported herein pave the way toward next generation MRI coils.

## MAIN TEXT

### Introduction

Magnetic resonance imaging (MRI) has emerged as a cornerstone of contemporary medicine, revolutionizing the approach to diagnosing and treating a diverse range of health conditions. Through its provision of non-invasive, high-resolution images of the internal human body (1), MRI significantly influences clinical practices, encompassing disease detection (2), treatment planning (3), and therapeutic monitoring (4). The foundational principles underpinning MRI are rooted in the phenomenon of nuclear magnetic resonance. During this process, nuclei within the human body respond to an externally transmitted radiofrequency (RF) excitation field (B1+), leading to a relaxation process wherein they release an RF field (B1-). This released signal contains crucial details regarding anatomical structure, tissue composition, and information about pathological alterations (5, 6). In modern clinical MRI, the stimulating field B1+ often originates from a substantial RF coil placed within the MRI bore, commonly known as the birdcage coil (BC). While the BC ensures a sufficiently homogeneous transmitting field, it encounters significant challenges when employed as a receive coil. Issues such as notable RF energy leakage and reduced sensitivity arise due to the distance between the coil and the patient (7). Consequently, in the context of signal acquisition, receive-only coils or phased array multi-channel receive-only coils are commonly employed alongside the BC (8-11). These auxiliary coils, strategically positioned in close proximity to the relevant anatomical area of the human body, play a pivotal role in clinical MRI. They effectively reduce noise levels and provide substantial signal-to-noise ratio (SNR) gain by reducing the patient-to-coil transmission distance (12).

For many years, receive coil arrays have dominated the MRI bedside for their heightened sensitivity in signal acquisition. Nevertheless, conventional MRI receive coils, while efficient, are renowned for their bulky, fixed, and rigid configurations (13-16), potentially leading to patient discomfort, bothersome positioning, and, in certain scenarios, compromised signal sensitivity. Additionally, to ensure a high SNR gain, these coil designs predominantly target specific areas of the body, necessitating the presence of separate coils for accommodating different anatomical sites. Imaging centers typically need 5-7 separate coils on hand targeting various anatomies, such as brain, spine, torso, shoulder, knee, foot/ankle, cardiac, pelvis, etc., which greatly adds to the overall hardware cost. To address these challenges, recent developments in MRI receive coils have gravitated towards more form-fitting, flexible designs. This evolution involves the integration of various conductor trace options for wearable MRI coils, such as conductive threads (17), conductive elastomers (18), electrotextiles (19), copper braids (20), liquid metal tubes (21), and screen-printed traces (22). These methods have proven their reliability as viable alternatives to traditional coils, providing enhanced patient comfort and comparable SNR by conforming more closely to average anatomy. Simultaneously, the next generation of receive-only MRI coils is poised to adopt a more modular approach, enhancing versatility to mitigate the current challenges of high costs and limited utility associated with MRI coils (23-26). Modular coil systems offer flexibility in customizing the field-of-view (FOV) coverage based on specific applications. This results in a singular coil array system which is well-suited for various uses, eliminating the need for expensive coils dedicated to individual anatomical regions. However, similar to conventional rigid coils, these flexible, form-fitting, and modular coils often incorporate a substantial number of non-magnetic electronic components, feed boards, cable traps, and adapters. This adds to the overall bulkiness and expense of the coil, necessitating careful handling during routine daily imaging procedures.

Coaxial cables, characterized by an inner conductor surrounded by a concentric conducting shield, with the two separated by a dielectric layer, have traditionally found application in various RF contexts. These include serving as feedlines connecting radio transmitters and receivers to antennas, network connections, and radio broadcasting, among other applications. Coaxial cables offer distinct advantages over alternative transmission lines, ensuring efficient and reliable signal transmission by minimizing power losses and providing inherent protection to signal integrity from external electromagnetic interference. Capitalizing on these unique electromagnetic properties, coaxial cables have recently found novel applications in wireless non-radiative power transfer (27-30). They exhibit remarkable magnetic responses by eliminating the electric dipole moment inherent in conventional helical coils commonly used for wireless power transfer (31,32). With respect to their application to MRI systems, the focus herein, recent advancements involve the utilization of coaxial cable in the development of surface coils (33). Examples include high-impedance gloves for hand imaging (34), wearable coils for breast imaging (35), and knee imaging (36, 37). These surface coils, based on coaxial cables, effectively isolate themselves electromagnetically from neighboring elements, achieving desired decoupling effects. However, it is noteworthy that these implementations of coaxial cables have typically been hardwired to a direct RF feed and are not operated wirelessly, which necessitates bothersome positioning and compromises patient comfort.

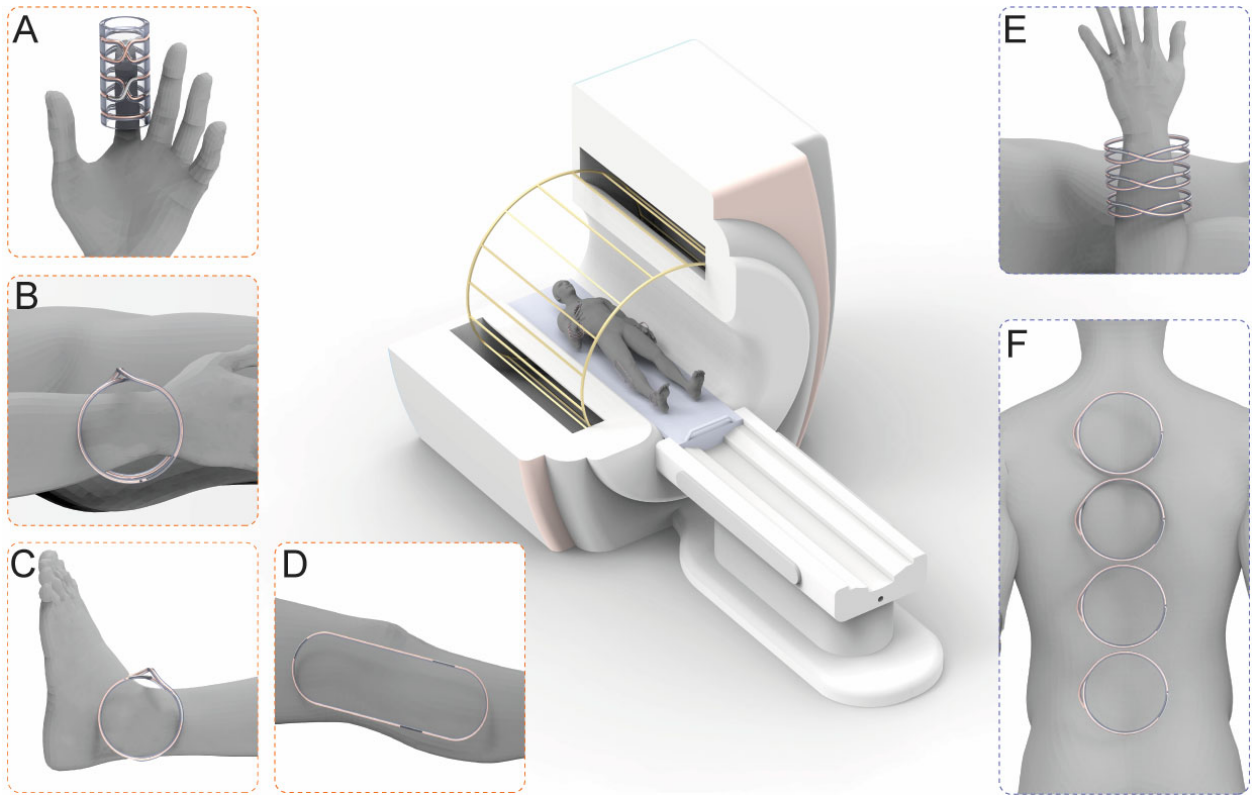

**Fig. 1. Application scenarios of the coaxially-shielded wireless coils in MRI systems.** (**A** to **D**) Form-fitting coil working independently for imaging finger (**A**), wrist (**B**), ankle (**C**), and knee (**D**). (**E** and **F**) Coaxial coils array (metamaterials) functioning collectively for imaging limb (**E**) and spine (**F**).

Harnessing the exceptional characteristics of coaxial cables in RF applications, we present a pioneering effort to design and fabricate wireless ultra-lightweight coaxial coils tailored for a 3.0 T MRI system. These coils, featuring multi-gap diode-loaded coaxially-shielded resonators, serve as additive components working in tandem with the BC, significantly amplifying MRI signals at a sub-wavelength scale. This achievement leads to an SNR comparable to, or even surpassing, that of commercial state-of-the-art receive coil arrays. Benefiting from their inherent flexibility and ease handling, the coaxially-shielded resonators are readily reconfigured to closely approximate small anatomical sites, such as the finger, wrist, and ankle, as illustrated in Figs. 1A~1D. Capitalizing on the coaxial cable's minimized dielectric loss and optimal form-fitting design, these coils exhibit heightened sensitivity in signal acquisition. This, coupled with enhanced patient comfort, a streamlined implementation process, high customizability, and significantly reduced costs, positions them as a groundbreaking advancement in the realm of MRI technology. Furthermore, to capture images of larger anatomical areas, such as the human knee or spine, we have developed coaxially-shielded metamaterials. These metamaterials consist of an array of coaxial resonators that are inductively coupled together to function collectively for signal acquisition, as illustrated in Figs. 1E and 1F. Metamaterials, constructed by assemblies of multiple judiciously designed structures at subwavelength scale, have emerged as powerful auxiliary devices to boost the imaging performance of MRI owing to their unique capacity for electromagnetic field confinement and enhancement (38-45). This innovative approach, facilitated by metamaterials, enables us to effectively expand coverage while preserving the high sensitivity characteristics of a single coaxial coil. In contrast to previously reported MRI metamaterials, the coaxially-shielded metamaterials provide a substantial SNR gain due to their remarkable magnetic response and the suppressed electric dipole moment found in conventional metamaterials. The metamaterial-enabled coaxial coil arrays provide a solution to wirelessly assemble coaxial coils for a customizable FOV coverage, leading to a modular coaxial coil array system suitable for multiple application scenarios, thereby eliminating the need for multiple expensive coils for each anatomical region.

## 2 Results

### 2.1 Coaxially-shielded circular resonators (CCRs)

To facilitate the design of the coaxial coils for MRI applications, we commenced by fabricating a coaxially-shielded circular resonator (CCR) with a diameter of 200/3 mm, aiming to achieve self-resonance close to the target Larmor frequency of a 3.0T MRI system. In the CCR design, the inner and outer conductors are separated by a dielectric layer, and gaps intersect both inner and outer conductors on opposite sides, as illustrated in Fig. 2A. These strategically positioned gaps at opposite ends of the conductors create a substantial structural capacitance between them, enabling the necessary resonance capacity despite the coil's small electrical size relative to the RF wavelength in an MRI system. Unlike resonators constructed with single conductive wires, CCRs do not incorporate lumped elements along the cable conductor. Instead, their resonance is solely dictated by their inherent geometric properties and total length, thus avoiding additional losses and promoting a high-quality factor in the context of MRI applications. The reflection spectrum of the CCR is illustrated in Fig. 2B, where the dips in the spectrum indicate the self-resonance frequency. At this self-resonance mode, the CCR generates a highly concentrated electric field in the narrow capacitive gaps between the inner and outer conductors. Simultaneously, there is a confined and enhanced magnetic field near the regions of the peak transient current, as depicted in Fig. 2C. With respect to MRI, when the frequency of the CCR's self-resonance mode approximates the working frequency of the MRI system, the CCR will be excited by the MRI RF signal (B1-) and generate a circulating current flowing along its conductive structures. In turn, the induced oscillating current will give rise to a magnetic field and dramatically enhance the incident B1- field, ultimately leading to gains in SNR.

The electric current within the CCR can be categorized into three distinct portions: the current along the surface of the inner conductor $I_{(in)}$, the current along the inner surface of the outer conductor $I_{(out)\_i}$, and the current along the outer surface of the outer conductor $I_{(out)\_o}$, as illustrated in Fig. 2C. Fig. 2D plots the simulated average magnitudes of these electric currents, offering a quantitative perspective of their relationship. $I_{(in)}$ and $I_{(out)\_i}$ have identical magnitudes but flow in opposite directions due to strong inductive coupling. These two currents start at 0 at the end of the inner conductor, reach their maximum value in the middle part of the conductor, and then return to 0, acting as the resonating current oscillating in a manner typical of a conventional resonator constructed by a single conductive wire. $I_{(out)\_i}$ and $I_{(out)\_o}$ are separated due to the skin effect at RF frequencies. At the position of the outer gap, the magnitudes of $I_{(out)\_i}$ and $I_{(out)\_o}$ are equal due to their interconnection between the inner and outer surfaces of the outer conductor. Leveraging the unique current profiles at the resonance mode, the inductive magnetic field is exclusively produced by $I_{(out)\_o}$ because the magnetic fields generated by $I_{(in)}$ and $I_{(out)\_i}$ cancel each other out. CCRs possess great resilience against frequency detuning due to load variations by confining the electromagnetic field carrying the signal exclusively to the capacitive gap between the inner and outer conductors. Compared to the conventional approach using copper wire or trace-based coil resonators, the primary advantage of CCRs lies in their predominantly magnetic response (Supplementary Text S1 and Fig. S1). The high degree of confinement of the electric field within the coaxial cable layers has significant practical implications in clinical MRI for suppressing the noise and offering a marked SNR enhancement by mitigating the undesired electric coupling to the sample. Next, we investigated the magnetic field in the vicinity of the CCR, and the magnetic field patterns on different cutting planes are shown in Fig. 2E. These enhanced magnetic field patterns highlight the CCR's potential to serve as a wireless receive coil for boosting MRI signals.

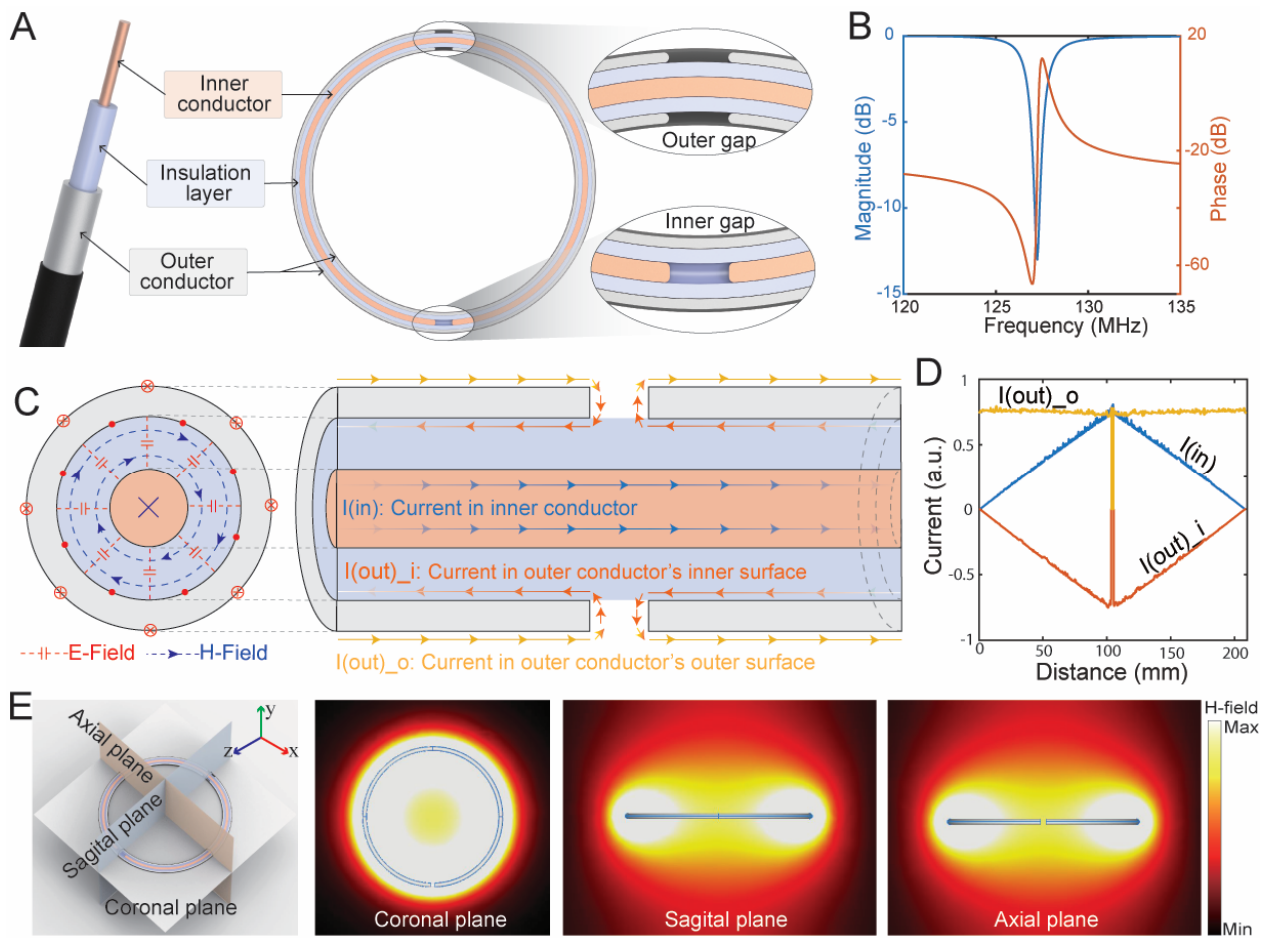

**Fig. 2. EM Characterizations of a CCR.** (**A**) Configuration of a CCR. (**B**) Reflection spectrum of a CCR. (**C**) Electric current profiles along inner and outer conductors' surfaces at resonance mode. (**D**) Electric currents' distribution. (**E**) Magnetic field pattern on coronal, sagittal, and axial cutting planes of a CCR at its resonance mode.

## 2.2 Towards frequency tunable, size customizable, and Self-adaptive CCRs

CCRs hold great potential for MRI applications owing to their distinctive resonating current profile and the resulting magnetic field distribution generate in their proximity. Despite its promise, the current implementation of the CCR shown in Fig. 2A faces practical limitations, hindering the realization of its full potential. Notably, these constraints include deviations in optimal resonance frequency, size limitations, and unavoidable interference with the RF transmission field (B1+) in MRI. To achieve optimal enhancement of the resonant magnetic field in MRI, it is crucial to align the working frequency of the CCR with that of the MRI system. While the resonance frequency of the CCR exhibits a certain degree of resilience against load variations, it remains susceptible to influences from the local environment and materials with varying permittivities. The potential for undesirable frequency shifts persists, particularly when the CCR is in proximity to patients with diverse body compositions, encompassing varying levels of water, fat, muscle, or bone during an MRI scan. Additionally, structural deformation, including the reconfigurations of the coils discussed below, can also impact the resonance frequency by altering the coupling coefficient between the electric currents along the outer conductor's surface (Supplementary Text S2 and Fig. S2). To overcome these challenges, we propose the incorporation of an innovative 'tuning sleeve' into the CCR to achieve resonance frequency tunability. The 'tuning sleeve' is a segment of conductive sheathing that encases the CCR and can selectively cover outer gaps by rotating along the CCR's outer surface, as illustrated in Fig. 3A1. While the majority of the electric field is confined to the narrow gap between the inner and outer conductors, a portion of the electric field extends to the region near the outer gap. Exploiting this exposed electric field, we can manipulate the electric field distribution in the outer gap area by rotating the tuning sleeve, as depicted in Fig. 3A2. This adjustment allows us to control the effective capacitance of the CCR, thereby tuning its resonance frequency. To validate this mechanism, we conducted measurements of the CCR's reflection spectrum while sweeping the rotating angle of the tuning sleeve from 0° to 30°. The extracted resonance mode frequencies, presented in Fig. 3A3, demonstrate a ~60 MHz frequency tuning range achievable through the introduction of the tuning sleeve.

The self-resonance frequency of a CCR depends on its coil diameter and the inherent geometric properties of the coaxial cables. Consequently, the coil design is restricted to a specific diameter to ensure that the self-resonance aligns with the desired Larmor frequency. However, in the realm of MRI applications, the optimal coil size is contingent upon the target FOV and penetration depth $p$. For a lossless circular coil, it is advisable to set the diameter equal to $2p/\sqrt{5}$ to maximize SNR at a depth $p$ within a uniformly conductive sample (46). Furthermore, to customize CCRs for specific anatomical sites and accommodate varying sizes, these CCRs will be bent along a 3D curved body contour and configured into specific shapes. To meet these design requirements, we propose introducing multiple gaps (inner/outer gaps, as indicated in Fig. 2A) to the CCR. This modification enhances the design flexibility of the coil diameter, allowing for customization based on specific application needs. In Fig. 3B1, a CCR configuration is depicted featuring three pairs of inner/outer gaps. The incorporation of multiple gaps in the CCR leads to a multi-segment electric current profile of $I_{(in)}$. This profile exhibits several distinct segments with varying current values, ranging from 0 to a maximum value and back to 0, as illustrated in the upper part of Fig. 3B2. In contrast, $I_{(out)\_o}$ maintains a uniformly distributed current profile, as shown in the lower part of Fig. 3B2. The introduction of multiple gaps exclusively influences the CCR's resonance frequency without impacting the oscillation strength at resonance. A quantitative analysis of the relationship between resonance frequency and CCR diameter, considering both single and multiple pairs of inner/outer gaps (Fig. 3B3), reveals a significant decrease in resonance frequency with an increase in the diameter of a CCR with a single gap. In contrast, CCRs with multiple gaps exhibit relatively stable resonance frequencies across a range of diameters. Minor frequency detuning issues observed in multi-gap CCRs due to variations in diameter can be effectively compensated for by employing a tuning sleeve. The incorporation of multiple gaps in CCR design overcomes size constraints, allowing for flexible diameter choices while maintaining the desired resonance frequency, thereby facilitating the customization of CCRs for specific anatomical sites, all while ensuring operation at their self-resonance frequency.

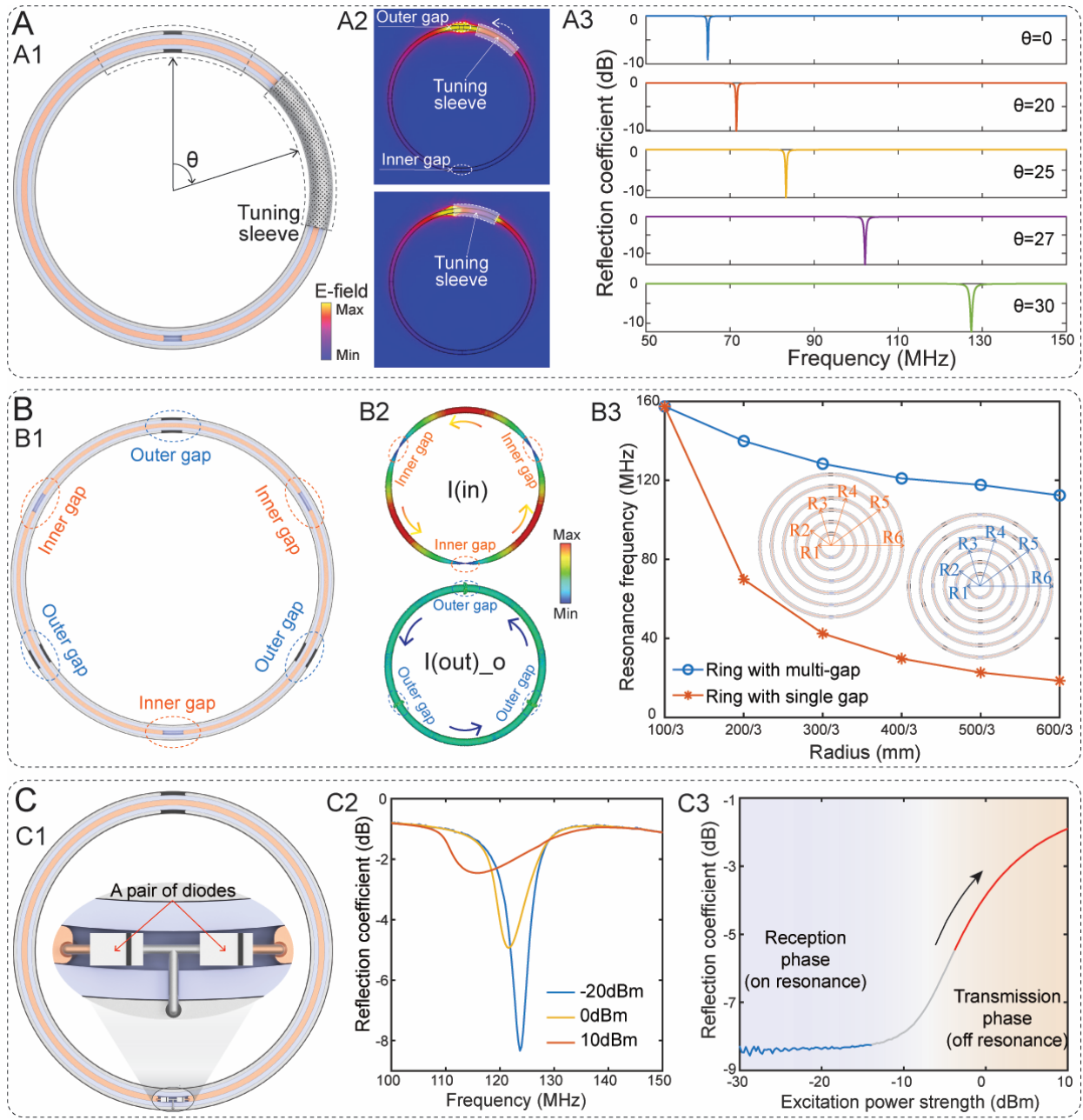

**Fig. 3. CCRs with frequency tunability, size customization, and self-adaptivity.** (**A**) Frequency tunability of CCRS. (A1) Illustration of CCR with tuning sleeve. (A2) Electric field patterns under different tuning angles. (A3) Reflection spectrum. (**B**) Multiple gaps enabled customizable size of CCRs. (B1) Configuration of CCR with multiple gaps. (B2) Electric current profiles. (B3) Comparison of CCRs' resonance frequency with multiple gaps and single gaps as diameter increases. (**C**) Self-adaptivity of CCR loaded with diodes. (C1) Configuration of CCR loaded with a pair of diodes. (C2) Nonlinear reflection spectrum. (C3) Reflection coefficient of CCR at its resonating state as a function of excitation strength.

Conventional LC resonators exhibit intrinsic linearity, resulting in magnetic field enhancement during both RF reception and transmission phases in MRI. While RF field B1-enhancement during reception improves SNR, amplifying RF field B1+ during transmission may causes issues such as unpredictable flip angle (FA) deviations, suboptimal performance, and potential safety concerns due to an increase in specific absorption rate (SAR). To address this, we introduced a pair of PIN diodes to each inner gap in the CCR, as shown in Fig. 3C1. To demonstrate its self-adaptive response to excitation power, we used a network analyzer coupled to a feeding loop with varying excitation strengths to characterize the resonance response, as plotted in Fig. 3C2. At an excitation power of -20 dBm, a deep dip in the reflection spectrum curve indicates strong oscillation amplitude at the resonance frequency, which yields a strong magnetic field enhancement in the vicinity of the CCR. As the excitation power increases, the CCR's resonant frequency shifts leftward, accompanied by a reduction in oscillation amplitude. At higher excitation powers, the diodes' rectifying effect acts as a driving voltage, increasing capacitance and contributing to the CCR's self-capacitance, ultimately decreasing the resonance frequency (47-49). Fig. 3C3 depicts the reflection coefficient at resonance states as a function of excitation power, illustrating variations in resonance strength across a range of excitation power levels, from low to high. In the case of MRI, the marked difference in power level between RF reception ($\mu$W) and transmission phases (kW) in MRI is sufficiently large to cause the CCR integrated with diodes to exhibit its self-adaptive response as a function of excitation power strength (50). This approach offers a promising method for designing self-adaptive CCRs for MRI, enabling an 'off' state during RF transmission and a desired 'on' state during RF reception.

## 2.3 Form-fitting CCRs for Targeting Small Anatomies.

Up to this point, we have independently realized the features of frequency tunability, size customization, and self-adaptivity in the CCR design, all of which are essential for making the CCR suitable for MRI applications. Through the integration of tuning sleeves, multiple gaps, and pairs of diodes, we successfully fabricated a series of CCRs with diverse diameters (Supplementary Text S3 and Fig. S3), and selected the CCR with diameter of 200mm depicted in Fig. 4A1 as the starting point for the following fabrication of form-fitting coils. To assess the CCR's electromagnetic properties, we conducted tests on its reflection spectrum, varying the incident wave's power levels and adjusting the tuning sleeve's angles, as illustrated in Fig. 4A2. At a specific tuning sleeve angle, with an increase in excitation power, the resonance mode of the CCR shifts to a lower frequency. Simultaneously, the oscillation amplitude diminishes, showcasing its self-adaptive response to excitation power. Furthermore, by adjusting the tuning angle from 0º to 60º, we achieve an ~30MHz tuning range in the resonance frequency. This range is sufficient to compensate for detuning effects during imaging, enabling an optimized frequency match between the CCR and the MRI system. Importantly, the integration of these features, i.e., tunability, size customization, and self-adaptivity, will not affect each other, ensuring that these features can work effectively in MRI.

The proximity of MRI coils to target anatomies is paramount for achieving high magnetic coupling, signal reception, depth of penetration, and image quality. To create form-fitting coaxial coils for MRI, capable of closely conforming to the size and shape of targeted anatomical sites, we reconfigured the CCRs by bending and shaping them along 3D curved body contours and curvatures. Initially, we transformed the circular CCR into a straight oval slot shape (Fig. 4B1), specifically designed for spine imaging. The form-fitting configuration of the CCR induces a redistribution of the magnetic field in the surrounding area of the coil. The magnetic field pattern at its resonance frequency on the cutting plane is illustrated in Fig. 4B2, showcasing its adaptability for spine imaging. By positioning the overall coil in close proximity to the spine, the thin strip configuration of the CCR maximally prevents the signal released from the spine from dissipating into the surrounding environment. Of course, the form-fitting spine coil is not restricted to spine imaging; it is suitable for any target anatomies with narrow and elongated shapes. Figs. 4B3 and 4B4 display photos of a person wearing the fabricated form-fitting spine coil for both arm and spine imaging. Besides the planar configuration, the CCR can be transformed into 3D snap-fit structures for imaging relatively small extremities, as depicted in Fig. 4C1. Geometrically, this form-fitting coil for small extremities allows the coil to be easily snapped or clicked onto the arm, wrist, or ankle without the need for additional tools or fasteners, providing an efficient means of positioning the coil during MRI scanning processes. On the electromagnetic side, when the form-fitting extremity coil is excited at its resonance frequency, the induced electric current in the top and bottom parts aligns in the same direction. This alignment provides a high degree of magnetic field enhancement through superimposition rather than cancelation of the magnetic fields arising from the induced currents. The simulated magnetic field pattern on its cutting plane is illustrated in Fig. 4C2, demonstrating a significant enhancement of the magnetic field inside the form-fitting coil. Figs. 4C3 and C4 further demonstrate the designed form-fitting extremity coil, showcasing its adaptability to different anatomies. Leveraging the CCRs' robustness against coil deformation and its cable-free design, the circular CCR can be dramatically twisted, wound, and shaped to conform to extremely small anatomical sites—a challenge with conventional rigid cabled MRI coils. While shaping CCRs from a circular ring into various 3D configurations may influence its resonance frequency, any frequency shift could be compensated for by the tuning sleeve. The form-fitting coil for finger imaging, illustrated in Fig. 4D1, demonstrates a super-strong field enhancement at its resonance frequency (Fig. 4D2), resulting in a high SNR gain when applied to MRI. Fig. 4D3 shows a photo illustrating a man wearing the coaxial coil on the finger, demonstrating its strong adaptability and flexibility. Beyond the configurations presented here, the CCR is highly customizable for any other anatomical sites.

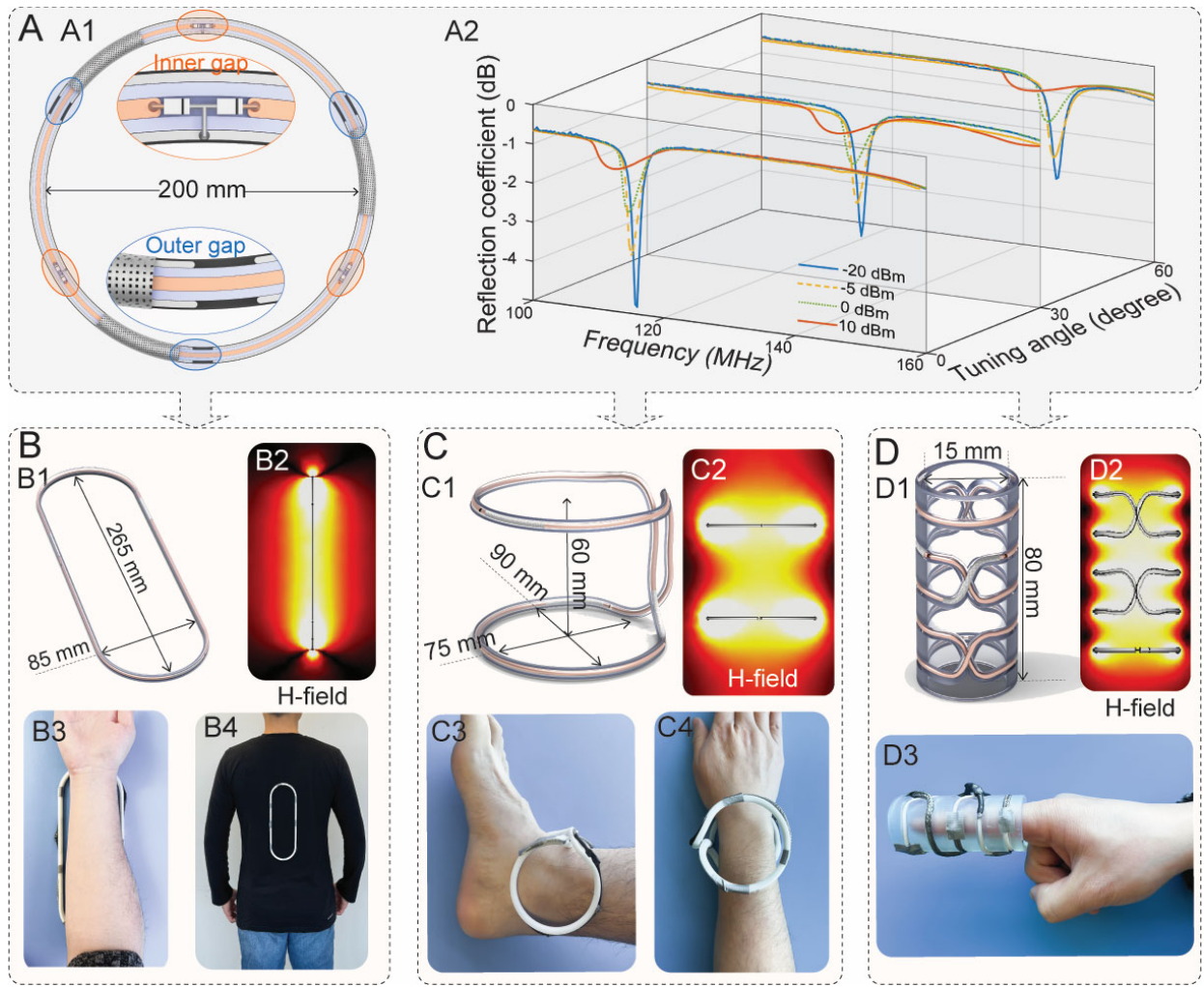

**Fig. 4. Form-fitting coils from the reconfigurations of CCR.** (**A**) Configuration of an CCR with 3-pairs of gaps (A1) and its reflection spectrum under different excitation power strength and various tuning angles (A2). (**B**) The configuration (B1) magnetic field pattern (B2) and application scenarios (B3 and B4) of the form-fitting CCR for spine imaging. (**C**) The configuration (C1), magnetic field pattern (C2), and application scenarios (C3 and C4) of the form-fitting CCR for small extremities imaging. (C) The configuration (C1), magnetic field pattern (C2), and application scenario (C3) of the form-fitting CCR for finger imaging.

## 2.4 Metamaterial-enabled CCR array for Targeting Large Anatomies.

The proposed CCR has potential in enhancing imaging capabilities for relatively small anatomical regions by shaping it into conformal structures. However, clinical MRI is a versatile imaging modality applied to visualize a wide range of larger anatomical parts, such as the human knee and spine. While the proposed form-fitting coaxial coils can adapt their working frequency by varying the number of gaps to accommodate different coil dimensions, thereby extending the coverage of a single coil to encompass larger anatomical regions, unavoidable compromises in coil sensitivity accompany broader coverage. This trade-off diminishes their advantages when compared to conventional receive coils. To address this limitation, we propose the development of metamaterials featuring an array of CCRs inductively coupled together to function collectively for signal acquisition. This approach allows for effective expansion of coverage while maintaining the high sensitivity characteristics of a single CCR.

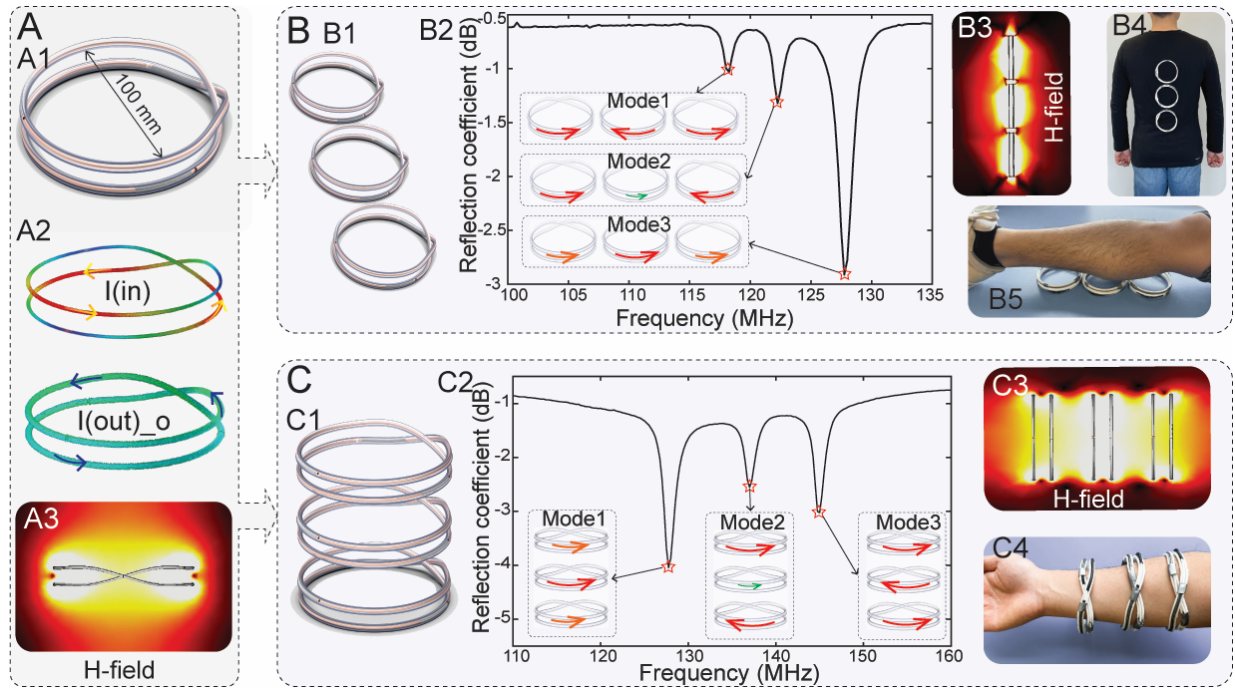

**Fig. 5.** (**A**) The configuration (A1), electric current profiles (A2), and magnetic field pattern (A3) of the 2-turn CCR with diameter of 100mm. (**B**) The configuration (B1), reflection spectrum (B2), magnetic field pattern (B3), and application scenarios (B4 and B5) of H-metamaterial. Inset of (b2): Electric current directions in each unit cell at three resonance modes. (**C**) The configuration (C1), reflection spectrum (C2), magnetic field pattern (C3), and application scenario (C4) of V-metamaterial. Inset of (C2): Electric current directions in each unit cell at three resonance modes.

The design of a metamaterial-enabled coaxial coil array begins with the reconfiguration of the CCR shown in Fig. 4A1. Initially, a CCR with a diameter of 200mm is reshaped into a 2-turn CCR with half the diameter (Fig. 5A1), aiming to strike a balance between high sensitivity and coverage area. The simulated resonating electric current along the inner and outer surfaces of the 2-turn CCR is plotted in Fig. 5A2. These current profiles result in a highly concentrated magnetic field at the central area of the coil, as demonstrated in Fig. 5A3. Next, the 2-turn CCRs are employed as building blocks to construct metamaterials for MRI applications. As illustrated in Fig. 5B1, the 2-turn CCRs are horizontally assembled into a 3-unit array to form the H-metamaterial. Once arranged in a metamaterial, the coupling effect between discrete coils must be considered, as it impacts the resonant mode of the coil array (51). The experimental reflection spectrum of this H-metamaterial, depicted in Fig. 5B2, reveals three distinct resonant modes, identified as dips on the plotted curves. The resonating current profiles in each unit corresponding to the three resonant modes are acquired through simulation, as illustrated in the inset figure of Fig. 5B2. For resonance modes 1 and 2, the direction of the electric current is different in each unit cell, resulting in a cancellation of the magnetic fields and a weakening of the MRI signal. In contrast, resonance mode 3, with the highest resonant frequency, exhibits identical electric current directions in each unit cell—referred to as the working mode for magnetic field enhancement (Supplementary Text S4, Fig. S4, and Table S1). In this working mode, the generated magnetic field presents a dramatic localization, leading to field confinement and enhancement (Fig. 5B3). The array design maintains high sensitivity and extends coverage through assembly and coupling, providing a large coverage without degrading performance. Figs. 5B4 and B5 illustrate application scenarios of a person wearing the H-metamaterial for spine and leg imaging. Beyond horizontal assembly, vertical stacking of the 2-turn coil forms the V-metamaterial for specific anatomies, as illustrated in Fig. 5C1 for arm imaging. Resonance mode analysis and reflection spectrum for the V-metamaterial are presented in Figs. 5C2 and C3, respectively (Supplementary Text S5, Fig. S5, and Table S2). Geometrically, the V-metamaterial is suitable for certain anatomies that can be encased by the V-metamaterial, as demonstrated in Fig. 5C4 with a person's arm. It is noteworthy that, based on coupled mode theory (51), the coupling coefficient between the units in the metamaterial can be manipulated by controlling their separation distance, adding another degree of freedom to tune their resonance frequency. The examples of the H-metamaterial and the V-metamaterial presented in Figs. 5B1 and C1, respectively, serve as demonstrations of how the coils can function collectively. The assembly of unit cells can be extended by employing more coaxial coils, and the dimension of the 2-turn coil is highly customizable due to the previously introduced multi-gap design. The combination of metamaterial technology with the great flexibility in the design of CCR provides enormous potential and possibilities for its use in MRI applications.

## 3 MRI validations

### 3.1 MRI validations for form-fitting CCRs with phantoms

In order to validate the SNR performance of the proposed coils when functioning with the BC in MRI, experimental validations with phantoms were conducted on a clinical 3T MRI system (Philips Healthcare). Various home-made phantoms were fabricated by filling 1% agarose gel into 3D printed modules to demonstrate the conformability of the coils. The MRI validations commenced with the form-fitting coils designed for spine imaging (shown in Fig. 4B). The phantom's configuration resembled a bottle shape with a diameter of 90 mm and a height of 280 mm, simulating the scenario of human spine imaging. The phantom was initially scanned by the BC alone through a gradient echo (GE) imaging sequence to assess the imaging power of the BC in the absence of the proposed coil, serving as a reference standard. Subsequently, the form-fitting coaxial spine coil was positioned below the phantom (as depicted in Fig. 6A1), and the system was scanned by the BC in the presence of the proposed coil. For comparison, the FlexCoverage Posterior coil array (Philips Healthcare) was employed as the control group. As the FlexCoverage Posterior coil array is integrated into the tabletop, the bottle phantom was placed directly on the bed without coil handling or positioning, as shown in Fig. 6A2. Using an identical imaging sequence, the corresponding SNR images scanned by BC only, BC combined with the coaxial spine coil, and the FlexCoverage posterior coil array are displayed with the same scale bar in Figs. 6A3, A4, and A5, respectively. For quantitative comparisons, the SNR enhancement ratio along the solid and dashed lines in Figs. 6A4 and A5 was extracted by normalizing to the mean value of SNR in Fig. 6A3. The results are plotted in Fig. 6A6, revealing a substantial enhancement in overall SNR, up to 10-fold compared to the acquisition by BC only. This indicates a significant boost in the imaging power of BC by integrating the coaxial spine coil into the MRI system. Comparing with the FlexCoverage posterior coil array, along the solid lines (in Figs. 6A4 and A5), the proposed coaxial coil provides comparable SNR performance. However, along the dashed lines (in Figs. 6A4 and A5) slightly shifting away from the central area, the proposed coaxial coil achieves a higher SNR by bringing the imaging site closer to the coil. Importantly, since the fixed FlexCoverage posterior coil array is embedded into the bed, its optimal SNR performance is limited to the bed's top surface. In contrast, the proposed coaxial coil's ultra-lightweight and cable-free design allows it to be attached to any anatomy of interest. As a result, the optimal SNR can be brought to the target area and breaks the limitation of FlexCoverage posterior coil array.

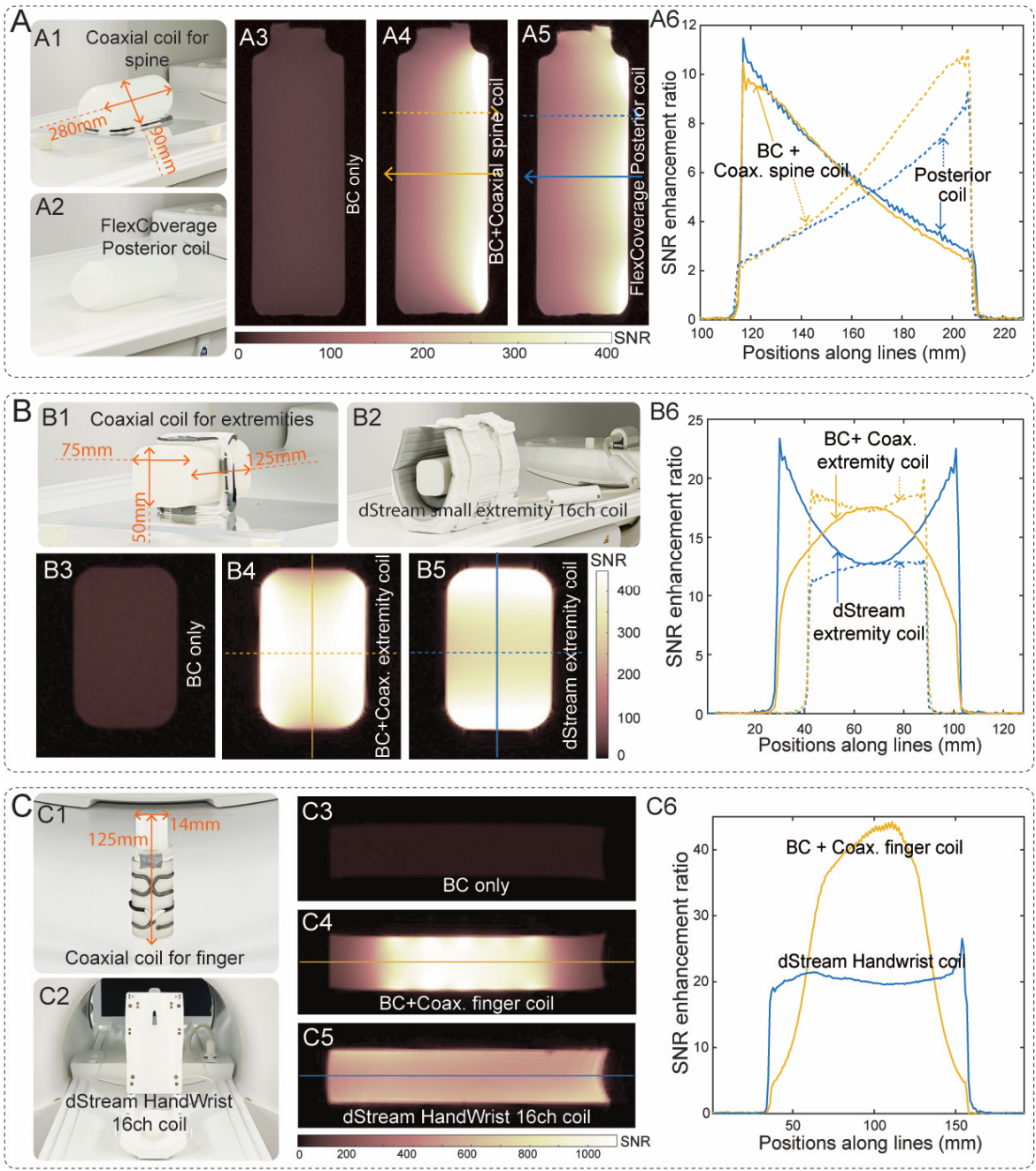

**Fig. 6. MRI validations with phantoms for form-fitting coaxial coils.** (**A**) Experimental setups for coaxial spine coil (A1), and Philips FlexCoverage Posterior coil array (A2). The SNR images by BC only (A3), BC combined with coaxial spine coil (A4), and posterior coil (A5). The SNR enhancement ratio normalized to BC only image (A6). (**B**) Experimental setups for coaxial extremity coil (B1) and Philips dStream small extremity 16ch coil (B2). The SNR images by BC only (B3), BC combined with coaxial extremity coil (B4), and 16ch extremity coil (B5). The SNR enhancement ratio (b6). (**C**) Experimental setups for coaxial finger coil (C1), and Philips dStream HandWrist 16ch coil (C2). The SNR images by BC only (C3), BC combined with coaxial finger coil (C4), and 16ch HandWrist coil (C5). The SNR enhancement ratio (C6).

Similar to the coaxial spine coil, the SNR performance of the coaxial extremity coil was also investigated in the MRI system. A 1% agarose gel phantom was configured in a rectangular prism shape with dimensions of length 125 mm, width 50 mm, and height 75 mm to mimic the small extremities of the human body. As the coaxial extremity coil is designed for small extremities imaging, the dStream small extremity 16ch coil (Philips Healthcare), a commercially available state-of-the-art surface coil targeting small extremities, was chosen as a counterpart to evaluate the performance of the proposed coaxial coil. The experimental setups are illustrated in Figs. 6B1 and B2, respectively. The corresponding SNR images on the axial plane, displayed with the same scale bar, are depicted in Figs. 6B3, B4, and B5 to represent the imaging power of BC only, BC integrated with the coaxial extremity coil, and the dStream small extremity 16ch coil, respectively. The normalized SNR enhancement ratio along the solid and dashed lines in Figs. 6B4 and B5 are plotted in Fig. 6B6. With the introduction of the coaxial extremity coil to the MRI system, the imaging power of the BC was boosted by ~17.5-fold. This degree of enhancement in the imaging power of BC by integrating with the coaxial extremity coil is stronger than that of the coaxial spine coil, resulting from the closer distance between the imaging site and the coil. Comparing with the dStream extremity 16ch coil, the proposed form-fitting coaxial extremity coil provides an overall better SNR performance. The SNR enhancement ratio, normalized to the BC-only image, is increased from ~12.7 to ~17.5 in the central region of the scanned image when switching from the dStream extremity 16ch coil to the coaxial extremity coil.

Replicating the experimental approach used for the coaxial spine and extremity coils, MRI validations were performed for the form-fitting coaxial finger coil. Initially, a small cylinder phantom with a diameter of 14 mm and a height of 125 mm was constructed to mimic the human finger. The state-of-the-art commercially available dStream HandWrist 16ch coil (Philips Healthcare), designed for finger imaging, was selected as a counterpart for fair comparisons. The experimental setups are illustrated in Figs. 6C1 and C2. The corresponding SNR images with the same scale bar, acquired by BC only, BC combined with the coaxial finger coil, and the dStream HandWrist 16ch coil, are depicted in Figs. 6C3 to C5. The imaging power of the BC is substantially boosted by introducing the coaxial finger coil to the MRI system, with a remarkable ~44-fold enhancement in the SNR. Even when compared with the dStream HandWrist 16ch coil, the coaxial finger coil outperforms it, providing up to a 2-fold enhancement in SNR value due to the ultimate close distance between the coil and the phantom, as shown by Fig. 6C6. In addition to the significant SNR improvement, the ultra-lightweight and cable-free design of the coaxial coil could enhance patient comfort and offer a simplified, faster workflow to increase productivity while maintaining excellent image quality. This is achieved by allowing for positioning freedom, especially in areas with previously challenging anatomies that are difficult to scan using conventional coils.

### 3.2 MRI validations for metamaterials with phantoms

In the following section, we will evaluate the SNR performance of the metamaterial-enabled coil arrays through MRI validation using the bottle phantom, as employed in the validation for the coaxial spine coil. The experimental setup for the H-metamaterial and the V-metamaterial is depicted in Figs. 7A and B, and the acquisitions based on these two setups are depicted in Figs. 7C and D, respectively. With the metamaterial design, both of these two arrays offer a larger coverage compared to a single coaxial coil, indicated by the larger area of the SNR enhancement patterns shown in these two SNR images. In contrast to the uniform pattern throughout the reference image by the BC only, the enhanced areas in the SNR images show similar patterns to the magnetic field pattern at resonance mode in Figs. 5B3 and C3, further supporting the direct relevance between the degree of magnetic field enhancement and eventual SNR performance during image acquisition. To conduct fair and thorough evaluations of the coil performance, we compared the SNR performance between the H-metamaterial-enabled coil array and the FlexCoverage Posterior coil, as their application scenarios have some degree of similarity. In parallel, we compared the V-metamaterial-enabled coil array with the dStream small extremity 16ch coil. As indicated by the SNR enhancement ratio shown in Fig. 7E, with an expanded coverage range, the SNR performance of the metamaterial-enabled coil arrays has not been compromised; instead, it shows a considerable improvement compared to advanced, clinical RF coils. The test results demonstrate the versatility of the CCRs in constructing customizable assemblies of coils, allowing for a FOV coverage tailored to various anatomies to be scanned.

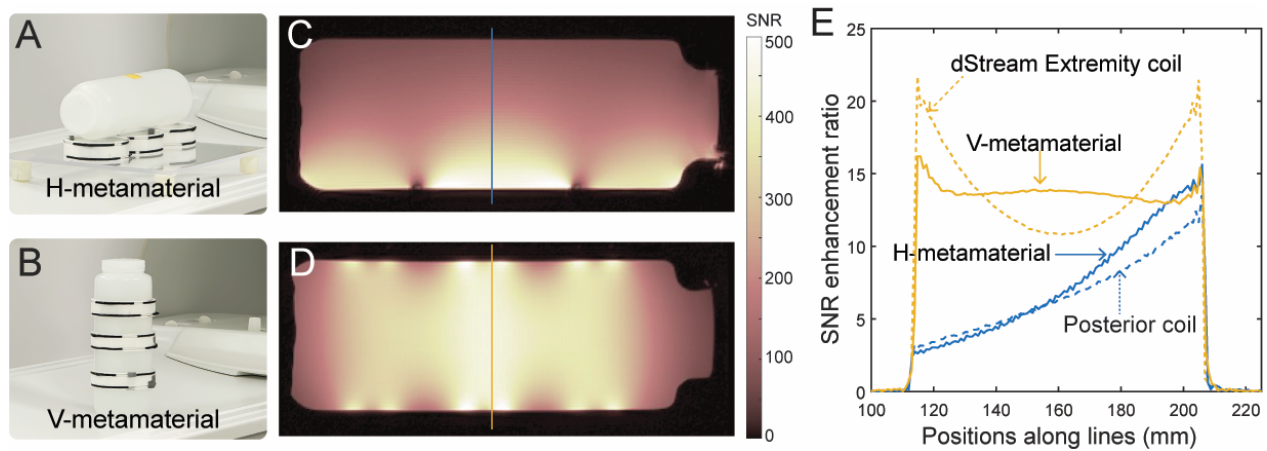

**Fig. 7. MRI validations with phantoms for metamaterials enabled coil array.** (**A** and **B**) Experimental setups for H-metamaterial (A) and V-metamaterial (B) enabled coaxial coil arrays. (**C** and **D**) SNR images correspond to experimental setup illustrated in (A) and (B). (**E**) Comparison of SNR enhancement ratios.

### 3.3 MRI validations with ex-vivo samples

To preliminarily demonstrate the performance of the coaxial coil in biomedically-relevant imaging, ex-vivo samples were employed for additional MRI validations. Besides the GE imaging sequence, an additional three mainstay pulse sequences commonly used in clinical MRI—T1-weighted turbo spin echo (T1w TSE), T2-weighted turbo spin echo (T2w TSE), and proton density-weighted turbo spin echo (PDw TSE)—were adopted to scan the ex-vivo samples, confirming the sequence compatibility of these coaxial coils. Firstly, a porcine leg was employed to evaluate the image quality. The images captured by BC only, FlexCoverage Posterior coil array, and BC combined with the coaxial spine coil are depicted in Fig. 8A. Using the same sample, we compared the acquisitions through the BC enhanced by the coaxial extremity coil and its counterpart, the dStream small extremity 16ch coil, as illustrated in Fig. 8B. Next, a chicken leg mimicking a human finger was adopted as the sample for comparing the image quality between the coaxial finger coil and its counterpart, the dStream HandWrist 16ch coil, as indicated in Fig. 8C. Lastly, the images of the porcine leg scanned by the H-metamaterial and V-metamaterial enabled coaxial coil arrays are depicted in Fig. 8D, validating their sensitivity when the coaxial coils function collectively to expand their coverage. Notably, through the comparison between the images acquired by BC in the absence and presence of the coaxial coil, the introduction of coaxial coils to the MRI system dramatically boosted the BC's imaging power, even with different tissues included in the leg (fat, muscle, bones, and bone marrow). Additionally, the results prove that the wireless coaxial coils featuring a self-adaptive property readily operate with a variety of clinical RF transmission pulse sequences widely used in MRI. When compared to the state-of-the-art multi-channel coil arrays, the coaxial coils offer comparable or even better image quality arising from their closely approximating configurations and inherit electromagnetic properties. Even though the signal intensity captured by a single coaxial coil presents a certain degree of decline with the expansion of its coverage, the metamaterial-enabled coil arrays provide a solution for targeting larger anatomies, realizing large coverages without compromising the coaxial coils' sensitivity. These images provide a preliminary demonstration of the potential of our technology to enhance imaging quality in complex anatomical settings.

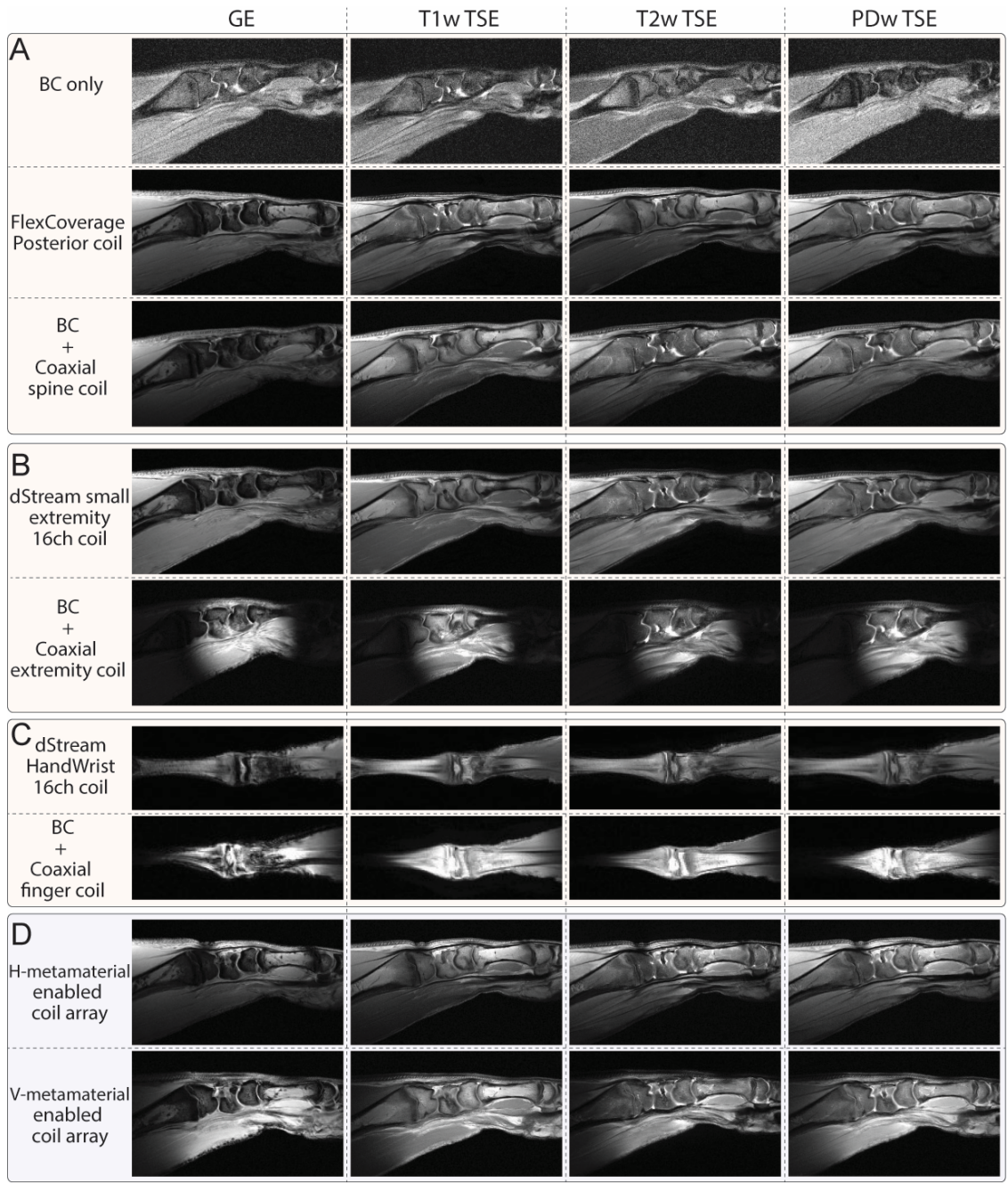

**Fig. 8. Comparisons of image quality captured by coaxial coils and their counterpart surface coil arrays.** (**A**) Comparisons of images captured by BC only, FlexCoverage Posterior coil, and BC enhanced by coaxial spine coil. (**B**) Comparisons between dStream small extremity 16ch coil and coaxial extremity coil. (**C**) Comparisons between dStream HandWrist 16ch coil and coaxial finger coil. (**D**) Images captured by H-metamaterial and V-metamaterial enabled coil arrays.

## 4. Discussion

This work introduces a novel paradigm for designing and constructing wireless, form-fitting, coaxially-shielded coils serving as auxiliary devices for the BC in a 3.0 Tesla MRI system, with the goal of achieving SNR and image quality comparable to or even better than state-of-the-art receive coil arrays, while enhancing patient comfort, simplifying implementation, providing high customizability, and significantly reducing costs. Leveraging the inherent structural flexibility and elimination of eddy currents, the proposed coils can function either as single closely form-fitting coils or collectively resonate as a metamaterial featuring a conformal array of coaxial coils. This allows them to adapt to a range of patient sizes, shapes, or challenging-to-image areas, offering high versatility and maintaining optimal SNR within the desired penetration depth for different use cases. The reported coaxial coils are applicable to various 3T MRI systems, irrespective of the manufacturer, and can be customized for MRI systems with different static magnetic field strengths, such as 1.5, 7.0, or 9.4 Tesla, conveniently facilitating their integration into different MRI systems. Since the superior properties of coaxial coils primarily arise from their internal structure, these coils are not limited to specific configurations and can be readily tailored into different shapes to conform to specific anatomies with distinct shapes and dimensions, exhibiting significant promise for widespread utilization within MRI systems (Supplementary Text S6 and Fig. S6). A limitation of note of the proposed coils is that they may not be compatible with parallel imaging due to their cable-free design. However, the proposed coils could provide a comparable or even better SNR than current commercially available surface coils at extremely low cost. The inexpensive, highly customizable coils have great potential of increasing the availability of lower cost MRI to society and finding many diverse applications throughout the MRI landscape. The investigation on coaxially-shielded coil resonators elucidates their governing physical mechanisms and provides a profound analysis of their performance in near-field enhancement. This offers a promising pathway for future developments of application-oriented electromagnetic devices, extending beyond MRI applications to scenarios involving radio frequency near-field powering or communication.

## 5. Materials and Methods

*Fabrication of Coaxial Coils:* The coaxially shielded coils were fabricated using commercially available non-magnetic coaxial cable (9849, Alpha Wire), loaded with a pair of PIN diodes (MADP-000235-10720T, MACOM Technology Solutions) at each outer gap. The dimensions of the inner conductor, insulation dielectric layer, outer conductor, and tuning sleeve are 0.48, 1.42, 1.93, and 3.15 mm, respectively. The materials employed include silver-plated copper for the inner conductor, silver-plated copper braid for the outer conductor, and tin-plated copper braid for the tuning sleeve. The diameter of the coaxial coils shown in Fig. 4A1 is 200 mm. The supportive structures facilitating the form-fitting coaxial coils were created through 3D printing using polylactic acid filament.

*Numerical Simulation:* Numerical simulations were conducted using the CST Microwave Studio Suite with the frequency domain solver. In the simulation model, the coil dimensions mirrored those of the fabricated samples detailed above.

*Bench test for EM characterization:* We employed a vector network analyzer (VNA, E5071C, Keysight Inc) with an inductive loop to excite these coaxial coils (Supplementary Fig. S7), in which the excitation power was set to -10 dBm to mimic the B1- field in MRI. In the characterization of the self-adaptivity to excitation strength, the tuning sleeve was adjusted to tune the resonance frequency of the metamaterial to ~127.7 MHz. The reflection spectra S11 were measured with a sweep in excitation power from -30 to 10 dBm.

*MRI validations with phantom:* All phantoms, depicted in Figs. 6 and 7, were created by pouring a 1% agarose solution into 3D-printed modules and left to cool for 24 hours until solidified into agarose gel. For all MRI experiments involving agarose gel phantoms, a gradient echo (GE) sequence was utilized. The sequence parameters included a repetition time (TR) and echo time (TE) of 100 and 4.6 ms, respectively. The acquisition (ACQ) voxel size was set at 1×1 mm, and the slice thickness was 5 mm. The flip angle (FA) was adjusted to 25°, representing the Ernst angle for 1% agarose gel. The SNR evaluation employed the two-image method (52), where one image of the phantom was obtained using gradient echo imaging, and a noise image was captured by deactivating the transmission RF coil (refer to these two images in Fig. S8, Support Information). The SNR of the phantom image was calculated as the ratio between the mean value of the phantom image magnitude and the standard deviation of the noise image.

*MRI validations with ex-vivo samples:* Detailed information about the experimental setup and sequence parameters can be found in Supplementary Fig. S9 and Table S3 of the Supporting Information. The ex-vivo samples of the porcine leg and chicken leg employed in this work were obtained from a local abattoir.

## Acknowledgments

This research was supported by the Rajen Kilachand Fund for Integrated Life Science and Engineering. The authors thank Dr. Yansong Zhao for technical assistance and valuable discussions during MRI experiments. The authors thank Boston University Photonics Center for technical support.

## Competing Interest Statement

The authors have filed a patent application on the work described herein, application No.: 16/002,458 and 16/443,126. Applicant: Trustees of Boston University. Inventors: Xin Zhang, Stephan Anderson, Guangwu Duan, and Xiaoguang Zhao. Status: Active.

## Supporting Information

Supporting Information is available.